\documentclass[a4paper,fleqn,usenatbib]{mnras}

\usepackage[T1]{fontenc}
\usepackage{ae,aecompl}

\usepackage{graphicx}	

\title[Four hot DOGs in the microwave]{Four hot DOGs in the microwave}

\author[S. Frey et al.]{S\'andor Frey$^{1}$\thanks{E-mail:
frey.sandor@fomi.hu}, Zsolt Paragi$^{2}$, Krisztina \'Eva Gab\'anyi$^{1,3}$ and Tao An$^{4,5}$\\
$^{1}$F\"OMI Satellite Geodetic Observatory, PO Box 585, H-1592 Budapest, Hungary\\
$^{2}$Joint Institute for VLBI ERIC, Postbus 2, 7990 AA Dwingeloo, the Netherlands\\
$^{3}$Konkoly Observatory, MTA Research Centre for Astronomy and Earth Sciences, PO Box 67, H-1525 Budapest, Hungary\\
$^{4}$Shanghai Astronomical Observatory, Chinese Academy of Sciences, 80 Nandan Road, 200030 Shanghai, People's Republic of China\\
$^{5}$Key Laboratory of Radio Astronomy, Chinese Academy of Sciences, 210008 Nanjing, People's Republic of China}

\begin{document}

\date{Accepted 2015 October 14. Received 2015 September 14; in original form 2015 August 1}

\pagerange{\pageref{firstpage}--\pageref{lastpage}} \pubyear{2015}

\maketitle

\label{firstpage}

\begin{abstract}
Hot dust-obscured galaxies (hot DOGs) are a rare class of hyperluminous infrared galaxies identified with the {\em Wide-field Infrared Survey Explorer (WISE)} satellite. The majority of them is at high redshifts ($z \sim 2-3$), at the peak epoch of star formation in the Universe. Infrared, optical, radio, and X-ray data suggest that hot DOGs contain heavily obscured, extremely luminous active galactic nuclei (AGN). This class may represent a short phase in the life of the galaxies, signifying the transition from starburst- to AGN-dominated phases. Hot DOGs are typically radio-quiet, but some of them show mJy-level emission in the radio (microwave) band. We observed four hot DOGs using the technique of very long baseline interferometry (VLBI). The 1.7-GHz observations with the European VLBI Network (EVN) revealed weak radio features in all sources. The radio is free from dust obscuration and, at such high redshifts, VLBI is sensitive only to compact structures that are characteristic of AGN activity. In two cases (WISE J0757+5113, WISE J1603+2745), the flux density of the VLBI-detected components is much smaller than the total flux density, suggesting that $\sim$70--90 per cent of the radio emission, while still dominated by AGN, originates from angular scales larger than probed by the EVN. The source WISE J1146+4129 appears a candidate compact symmetric object, and WISE J1814+3412 shows a 5.1-kpc double structure, reminiscent of hot spots in a medium-sized symmetric object. Our observations support that AGN residing in hot DOGs may be genuine young radio sources where starburst and AGN activities coexist.     
\end{abstract}

\begin{keywords}
galaxies: active -- galaxies: high-redshift -- galaxies: starburst -- galaxies: individual (WISE J0757+5113, WISE J1146+4129, WISE J1603+2745, WISE J1814+3412) -- radio continuum: galaxies -- techniques: interferometric.
\end{keywords}

\section{Introduction}

One of the main scientific goals of the {\em Wide-field Infrared Survey Explorer (WISE)} space telescope \citep{Wrig10} was to find the most luminous galaxies in the Universe. {\em WISE} was sensitive in four infrared bands at wavelengths 3.4, 4.6, 12, and 22 $\mu$m (denoted by $W1$, $W2$, $W3$, and $W4$, respectively) and made an all-sky survey. Among the catalogued sources, $W1W2$-dropouts are heavily obscured galaxies, very faint or even undetected in the $W1$ and $W2$ bands, but have significant flux density in the longer-wavelength $W3$ and $W4$ bands \citep{Wu12,Eise12}. The dropouts have Vega magnitudes and colours as $W1>17.4$, and either $W4<7.7$ and $W2-W4>8.2$, or $W3<10.6$ and $W2-W3>5.3$. The detailed selection criteria are given by \citet{Eise12}. These objects are hyperluminous infrared galaxies \citep[HyLIRGs, $L_{\rm IR} > 10^{13} L_{\odot}$,][]{Rowa00,Eise12}. At redshifts of $z \sim 2-3$ corresponding to the cosmological epoch of peak star formation \citep[e.g.][]{Heav04}, the near-infrared obscuration in the rest frame of the galaxies is expected to fall in the $W1$ and $W2$ bands. On the other hand, the strong reradiated emission of the hot obscuring dust is detected at 12 and 22 $\mu$m. Spectroscopic observations of more than 100 $W1W2$-dropout galaxies \citep{Eise12,Brid13} confirmed that the majority of these galaxies are indeed at $z>1.6$, with the bulk of the sources at $z \sim 2-3$. The all-sky sample of $W1W2$-dropouts contains nearly 1000 objects. Their surface density ($\la 0.03$\,deg$^{-2}$) is more than 5 orders of magnitude smaller than that of {\em WISE} sources in general \citep{Eise12}\footnote{The {\em WISE} all-sky data release source catalogue from which the hot DOGs were selected is available at {\tt http://wise2.ipac.caltech.edu/docs/release/allsky} and contains nearly 564 million objects, corresponding to $\approx 1.4 \times 10^4$\,deg$^{-2}$ average surface density.}. High-redshift $W1W2$-dropouts are coined as hot DOGs \citep[hot dust-obscured galaxies,][]{Wu12}.

The optically faint dust-obscured galaxies (DOGs) selected with {\em Spitzer} have large ($\ga1000$) mid-infrared (24\,$\mu$m) to optical ($R$-band) flux density ratios \citep{Dey08}. They are thought to be in transition between the starburst-dominated and AGN-dominated phases in the process of a major galaxy merger. In this scenario, submillimetre galaxies \citep[SMGs,][]{Blai02} evolve into DOGs, then (optical) quasars, and finally ``red and dead'' elliptical galaxies \citep[e.g.][]{Nara10,Buss11}. Based on {\em Herschel} data, \citet{Rigu15} showed that faint DOGs (with 8-$\mu$m luminosity below $10^{12}$\,$L_\odot$) are dominated by star formation, while the AGN activity is typical for brighter DOGs. Most $W1W2$-dropout galaxies can also be classified as DOGs. Their redshift distribution is similar but the luminosities of $W1W2$-dropouts are typically an order of magnitude higher than those of ``normal'' DOGs \citep{Wu12}. Hot DOGs can be considered a $\sim$10\,000 time rarer extreme subclass of DOGs, characterised by very high dust obscuration and hotter dust temperatures. Because only very few hot DOGs are found in the all-sky survey, they likely represent a brief transitional phase of galaxy evolution at the end of the DOG phase \citep{Wu12}. 

Since their discovery, hot DOGs have become the subject of intense research at different wavebeands of submillimetre \citep[e.g.][]{Wu14,Jone14}, infrared \citep[e.g.][]{Asse15,Tsai15}, and X-rays \citep{Ster14,Pico15}. The prototype of this class, the first such a source investigated in detail with optical, infrared, submillimetre and radio imaging, and optical spectroscopy was WISE J1814+3412 \citep{Eise12}. Its radio emission, exceeding the amount expected from the far infrared--radio correlation \citep[e.g.][]{Cond92} {by a factor of $\sim$10 \citep{Eise12}}, is considered as the strongest evidence for the presence of an active galactic nucleus (AGN) in the system. For comparison, the ratio between the far infrared and radio emission in WISE J1814+3412 is similar to that of the radio excess sources studied with {\em Herschel} in the GOODS-North field by \citet{DelM13}. The radio emission originating from an AGN can penetrate through the dusty environment in the host galaxy. According to \citet{Eise12}, the observations of WISE J1814+3412 can be well modeled with a spectral energy distribution (SED) combined from an obscured AGN, a starburst, and an evolved stellar component. The star formation rate (SFR) is estimated to be $\sim$300~$M_{\odot}\,{\rm yr}^{-1}$ \citep{Eise12}. 

Submillimetre and infrared observations of a sample of 26 hot DOGs that were among the first such objects spectroscopically identified at high redshifts ($z \ga 2$) indicate that the SEDs are dominated by hot (60--120~K) dust emission generated by powerful AGN, but multiple components with different temperatures are also present \citep{Wu12}. In a follow-up project, \citet{Wu14} observed three hot DOGs drawn from the above sample with interferometric mm-wavelength observations using the Combined Array for Research in Millimeter-wave Astronomy (CARMA) and the Submillimeter Array (SMA). This pilot study with $\sim$1$^{\prime\prime}$ resolution led to the continuum detection of two unresolved sources out of the targeted three. This allowed the authors to estimate the upper limits of molecular gas and cold dust contents in these galaxies. Further submillimetre data of a partly overlapping sample of 10 hot DOGs confirmed that the SEDs cannot be well fitted by AGN templates alone, without considering extra dust extinction \citep{Jone14}. The first X-ray spectrum of a hot DOG (WISE J1835+4355) also indicates the presence of a heavily obscured AGN accompanied by extreme star formation \citep{Pico15}.   

The {\em WISE}-selected dusty Ly$\alpha$ emitters, a class that largely overlaps with hot DOGs at high redshifts, are typically radio-quiet \citep{Brid13}. However, some of the known hot DOGs with a yet unknown fraction do show mJy-level radio emission at 1.4~GHz. The radio emission provides us with an intriguing opportunity to directly confirm AGN-related radio emission in these sources since the dust in the galaxies is transparent at GHz frequencies. Upon detection, the uniquely high angular resolution achievable with the technique of very long baseline interferometry (VLBI) guarantees that the emitting region is compact. Moreover, if a radio source is detected with VLBI from a sufficiently large distance (at redshifts well above $z$$\sim$0.1), then the corresponding luminosity clearly exceeds that of the brightest supernovae and/or supernova remnant complexes \citep{Alex12,Midd13}. Therefore VLBI offers a sharp and unobscured view of the compact AGN-related structures if they are indeed present in radio-emitting hot DOGs.

\citet{Lons15a} studied heavily obscured hyperluminous infrared quasars selected via extremely steep mid-infrared (MIR) spectra obtained by {\em WISE}. Unlike for the samples mentioned above \citep{Eise12,Wu12,Brid13}, the list was cross-matched with 20-cm radio catalogues to select sources that show arcsec-scale compact radio emission. These sources share key properties of radio-emitting hot DOGs, such as similar SED, high luminosity and redshift. \citet{Lons15a} found that the MIR emission in these sources is dominated by AGN, but the SFR is loosely constrained, allowing values from zero up to a few thousand $M_{\odot}\,{\rm yr}^{-1}$. 

Here we report on our VLBI observations of four hot DOGs with mJy-level total emission in the cm--dm microwave bands. These observations were performed with the European VLBI Network (EVN) at 1.7~GHz on 2014 February 21--22. The sample selection is described in Sect.~\ref{sample}, the observations are detailed in Sect.~\ref{observations}, and the results are presented and discussed in Sect.~\ref{results}. To calculate linear sizes and luminosities \citep{Wrig06}, we assume a flat cosmological model with $H_{\rm 0}$=70~km~s$^{-1}$~Mpc$^{-1}$, $\Omega_{\rm m}$=0.3, and $\Omega_{\Lambda}=$0.7 in this paper.

\section{Sample Selection}
\label{sample}

\begin{table*}
  \centering 
  \caption[]{Parameters of the four hot DOGs observed with the EVN.}
  \label{sources}
\begin{tabular}{ccccccc}        
\hline                 
Source name     & Redshift        & \multicolumn{4}{c}{{\em WISE} flux densities$^{\rm c}$ (mJy)} & Total 1.44-GHz radio  \\
                &                 & $W1$ & $W2$ & $W3$ & $W4$                                     & flux density (mJy)    \\
\hline                       
WISE J0757+5113 & 2.227$^{\rm a}$ &   0.02  & $<0.04$ &  1.46 &  9.31 & 3.57$\pm$0.15$^{\rm d}$     \\
WISE J1146+4129 & 1.772$^{\rm a}$ &   0.03  &   0.06  &  3.90 & 20.35 & 4.39$\pm$0.13$^{\rm d}$     \\
WISE J1603+2745 & 2.633$^{\rm a}$ & $<0.02$ & $<0.04$ &  3.15 &  9.53 & 2.28$\pm$0.14$^{\rm d}$     \\
WISE J1814+3412 & 2.452$^{\rm b}$ & $<0.01$ &   0.02  &  1.86 & 14.38 & 1.43$\pm$0.18$^{\rm e}$     \\
\hline   
\end{tabular}
\\
$^{\rm a}$ \citet{Wu12}; 
$^{\rm b}$ \citet{Eise12};
$^{\rm c}$ calculated from Vega magnitudes according to \citet{Cutr12};
$^{\rm d}$ from the Faint Images of the Radio Sky at Twenty-cm (FIRST) survey \citep{Whit97};
$^{\rm e}$ estimated from the 4.49- and 7.93-GHz flux densities and the spectral index reported by \citet{Eise12}
\end{table*}

We selected four hot DOGs from the 26-element sample of \citet{Wu12}. These were the only mJy-level radio sources among hot DOGs known at the time of the project planning (Table~\ref{sources}). Three of them (WISE J0757+5113, WISE J1146+4129, WISE J1603+2745) are listed in the Very Large Array (VLA) Faint Images of the Radio Sky at Twenty-cm (FIRST) survey catalogue\footnote{\tt{http://sundog.stsci.edu}} \citep{Whit97} as unresolved ($<5^{\prime\prime}$) objects, with integral flux densities of $\sim2-4$~mJy. The detection threshold of the FIRST survey is $\sim$1\,mJy which is therefore the lower flux density limit for our selection of radio-emitting hot DOGs. From the full list of \citet{Wu12}, there are 9 non-detections in FIRST, implying sub-mJy flux densities at 1.4~GHz. The rest of the hot DOGs, further 14 objects from the 26-element \citet{Wu12} sample, lie outside the FIRST sky coverage. However, one of them, the fourth hot DOG in our radio sample (WISE J1814+3412) was observed and detected as an unresolved source ($<9^{\prime\prime}$) with the Karl G. Jansky VLA at two different frequencies, 4.5~GHz and 7.9~GHz \citep{Eise12}. From the measured spectral index, the extrapolated total 1.4-GHz flux density of WISE J1814+3412 is 1.4~mJy. Based on the total of 12 sources (9 non-detections and 3 detections) that are covered by FIRST, the radio-emitting fraction of hot DOGs is estimated as $\ga$25 per cent.

\section{Observations and Data Reduction}
\label{observations}

The EVN observations started on 2014 February 21, using a network of seven radio telescopes in Europe (Effelsberg in Germany, the Jodrell Bank Lovell Telescope in the UK, Medicina and Noto in Italy, Onsala in Sweden, Toru\'n in Poland, and the Westerbork Synthesis Radio Telescope in the Netherlands) and one in China (Sheshan). The experiment lasted for 14\,h. The observations were performed using the electronic VLBI (e-VLBI) technique \citep{Szom08}. This is different from the traditional VLBI where the data are recorded on magnetic media (disks) at each individual radio telescope and later played back at a central correlator facility to achieve the interference. In e-VLBI, the signals from the telescopes are streamed from the remote sites to the data processor over optical fibre network. The correlation of the data was done in real time with 1-s integration using the EVN software correlator \citep[SFXC,][]{Keim15} at the Joint Institute for VLBI in Europe (JIVE) in Dwingeloo, the Netherlands. The data were collected at 1024~Mbit~s$^{-1}$ rate with two circular polarizations, eight basebands per polarization, each with thirty-two 500-kHz spectral channels. This resulted in a total bandwidth of 128~MHz in both left and right circular polarizations, using 2-bit sampling. The central frequency was 1.659\,GHz.

The four targeted hot DOGs are too weak for fringe-fitting \citep{Schw83} if observed within the atmosphetic coherence time (typically $\sim$10\,min at this frequency). Therefore they were observed in phase-reference mode, to secure signal-to-noise ratios sufficient for detection by increasing the total coherent integration time spent on the targets. Phase-referencing is performed by regularly nodding the radio telescope pointing direction between the target source and a nearby bright, compact reference source \citep[e.g.][]{Beas95}. The delay, delay rate, and phase solutions derived for the phase-reference calibrators were interpolated and applied to the respective targets within the cycle time of $\sim$5\,min. Each target source was observed for about 3.5\,min within a cycle, accumulating $\sim$50\,min on-source integration time for WISE J0757+5113, and $\sim$130\,min for the other three targets, respectively. The target--reference pairs and their angular separations were the following: WISE J0757+5113 and J0756+5151 ($0\fdg63$), WISE J1146+4129 and J1146+3958 ($1\fdg52$), WISE J1603+2745 and J1606+2717 ($0\fdg82$), WISE J1814+3412 and J1826+3431 ($2\fdg64$). Additional fringe-finder and calibrator sources (J0824+5552, J1606+3124, J1734+3857) were also observed during the experiment.

The VLBI data were caibrated using the NRAO Astronomical Image Processing System\footnote{\tt{http://www.aips.nrao.edu}} \citep[{\sc AIPS}, e.g.][]{Diam95}. The visibility amplitudes were calibrated using antenna gains and system temperatures measured at all of the telescopes but Jodrell Bank, where nominal system temperatures were applied. Fringe-fitting \citep{Schw83} was performed for the phase-reference and other calibrator sources using 3-min solution intervals. These data were exported to the Caltech {\sc Difmap} package \citep{Shep94} where each calibrator source was imaged using the conventional hybrid mapping procedure involving several iterations of {\sc CLEAN}ing \citep{Hogb74} and phase (then amplitude) self-calibration. This provided images and brightness distribution models for the calibrators. Overall antenna gain correction factors (up to $\sim$20 per cent) were also determined in {\sc Difmap} and the average values were applied to the visibility amplitudes in {\sc AIPS} if exceeded 5 per cent.

Fringe-fitting was then repeated for the phase-reference calibrators, now taking their {\sc CLEAN} component model into account, to compensate for small residual phases resulting from their non-pointlike structure. The solutions obtained were interpolated and applied to the respective target source data. The calibrated and phase-referenced visibility data of WISE J0757+5113, WISE J1146+4129, WISE J1603+2745, and WISE J1814+3412 were also exported to {\sc Difmap} for imaging. 

\begin{figure*}
\begin{minipage}{85mm}
\centering
  \includegraphics[bb=70 75 524 720, width=60mm, angle=270, clip=]{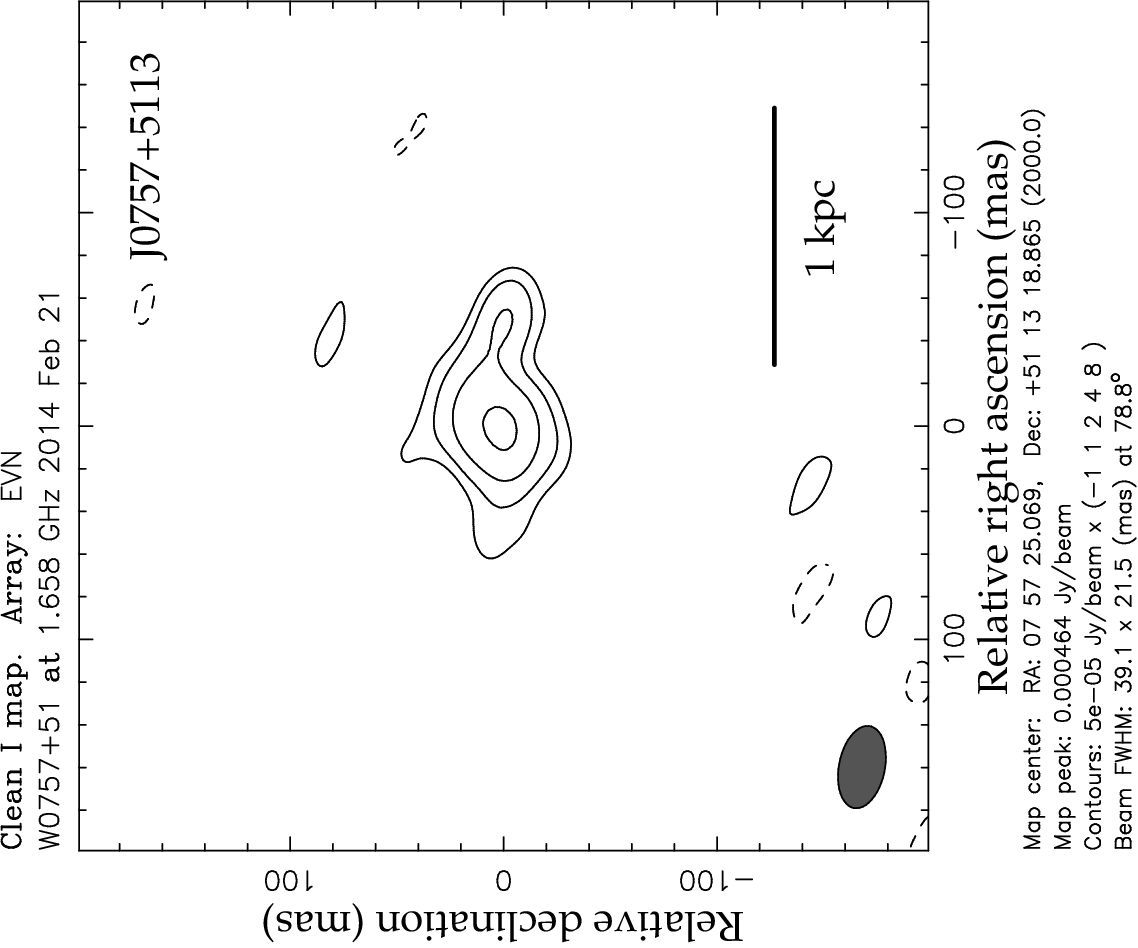}
\end{minipage}
\begin{minipage}{85mm}
\centering
  \includegraphics[bb=70 75 524 720, width=60mm, angle=270, clip=]{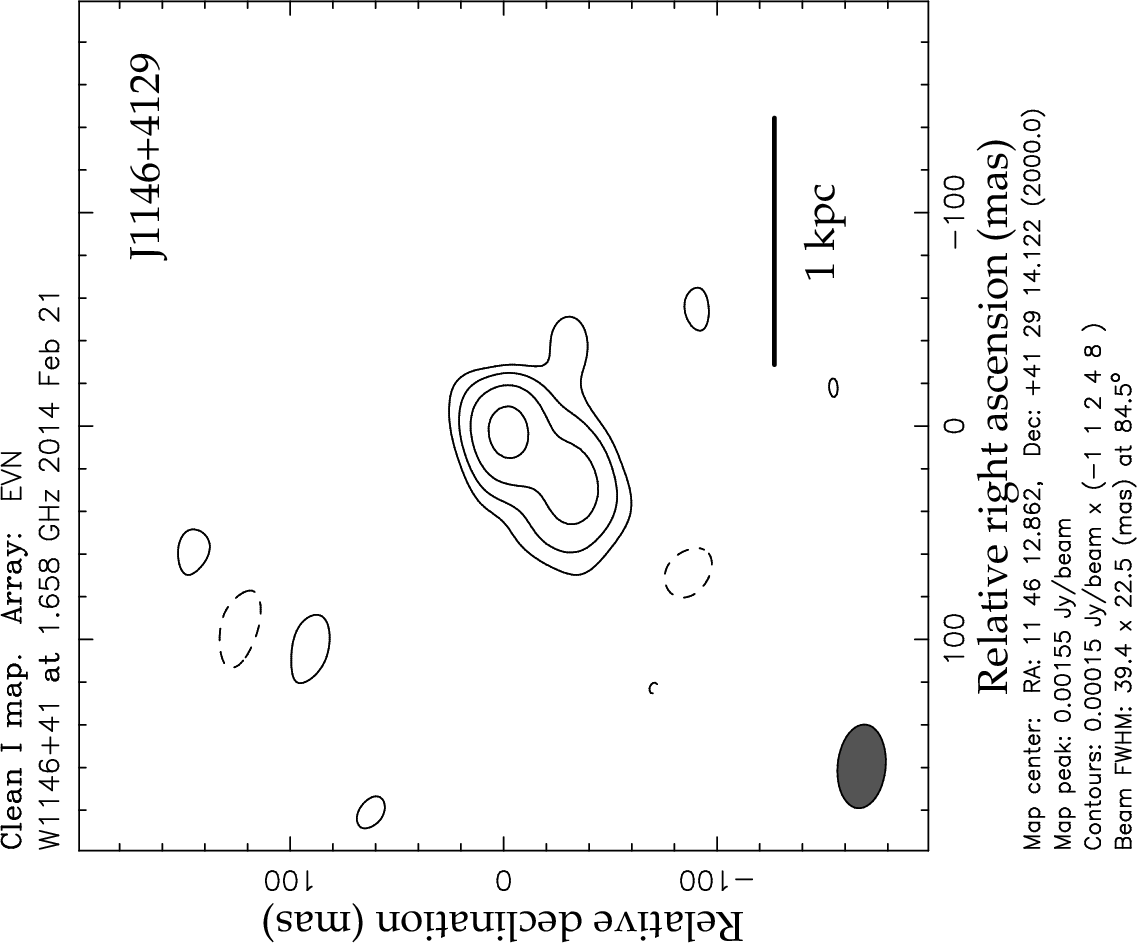}
\end{minipage}
\begin{minipage}{85mm}
\centering
  \includegraphics[bb=70 75 524 720, width=60mm, angle=270, clip=]{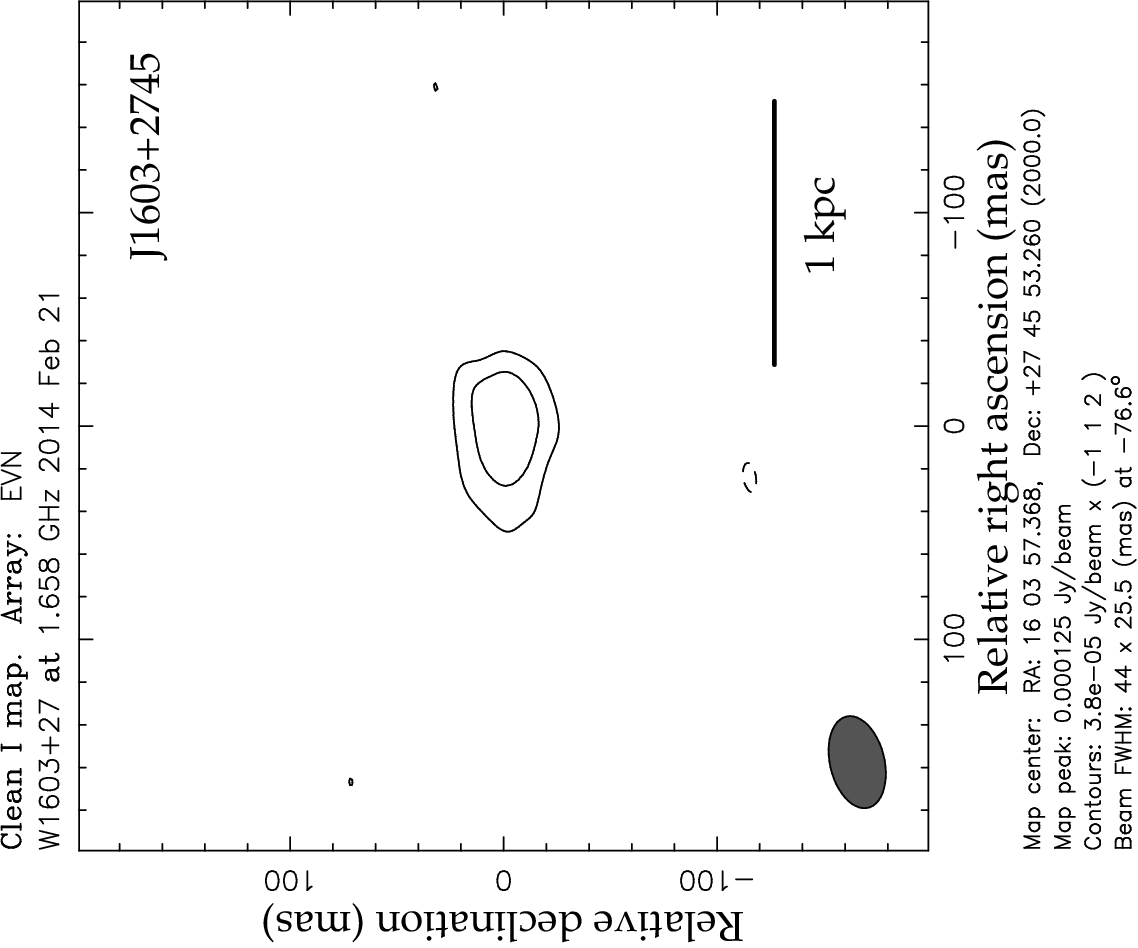}
\end{minipage}
\begin{minipage}{85mm}
\centering
  \includegraphics[bb=70 125 524 671, width=60mm, angle=270, clip=]{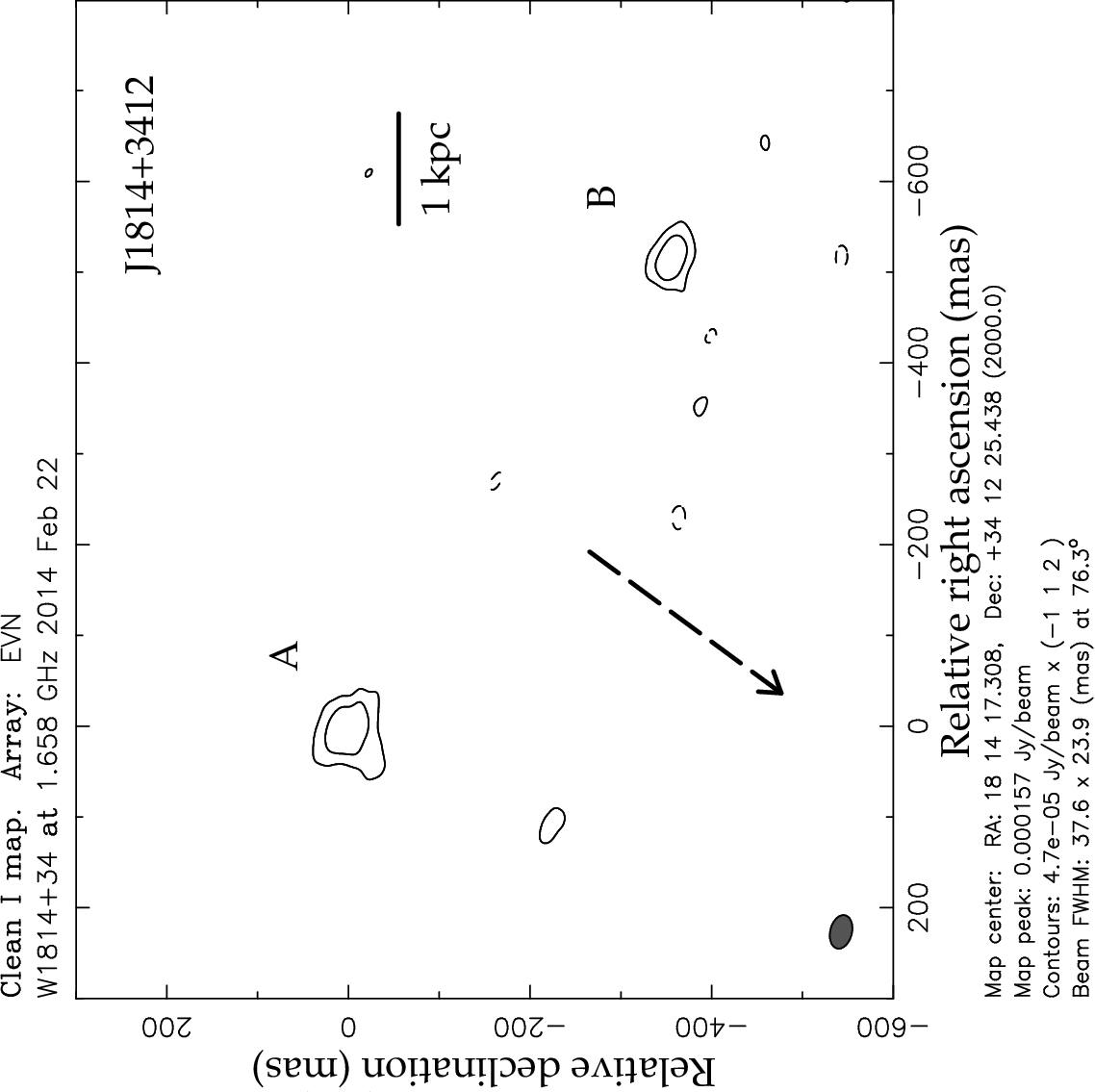}
\end{minipage}
  \caption{
The naturally-weighted 1.7-GHz EVN images of WISE J0757+5113, WISE J1146+4129, WISE J1603+2745, and WISE J1814+3412. The image parameters are listed in Table~\ref{imagepar}. The lowest contours are drawn at $\sim$4$\sigma$ image noise level. Further positive contour levels increase by a factor of 2. The Gaussian restoring beams are indicated with filled ellipses (full width at half maximum, FWHM) in the lower-left corners. The coordinates are related to the brightness peaks. The absolute source positions are given in Table~\ref{sources-derived}. In case of WISE J1814+3412, the position angle of the Ly$\alpha$ blob extending to $>30$~kpc found by \citet{Eise12} from optical spectroscopic observations is indicated with a dashed line.}
  \label{images}
\end{figure*}

The total intensity images at 1.7~GHz are presented in Fig.~\ref{images}. The sources were all heavily resolved, i.e. their sizes were larger than the resolution limit of the interferometer array, and thus the longest intercontinental baselines had little contribution to the signal. We therefore omitted data from Sheshan when producing the final images. The image parameters are given in Table~\ref{imagepar}. The images were restored in {\sc Difmap} after iteratively fitting circular Gaussian brightness distribution models to the interferometric visibility data, until $\sim$6$\sigma$ noise level was reached in the residual image. Finally, for smoothing the residual features smaller than the restoring beam, a deep {\sc CLEAN} procedure with 1000 iterations and 0.01 loop gain was applied. Natural weighting (weighting the visibility data points by the inverse square of amplitude errors) was employed to minimize the image noise.    

\begin{table*}
  \centering 
  \caption[]{Parameters of the images in Fig.~\ref{images}. The restoring beam major axis position angle (P.A.) is measured from north through east.}
  \label{imagepar}
\begin{tabular}{cccccc}        
\hline                 
Source name     & Peak brightness    & Lowest contour     & \multicolumn{3}{c}{Restoring beam} \\
  &                    &                    & major axis  & minor axis & P.A.  \\
  & ($\mu$Jy~beam$^{-1}$) & ($\mu$Jy~beam$^{-1}$) & (mas)       & (mas)      & ($\degr$) \\
\hline                       
WISE J0757+5113 & 464                & 50                 &  39.1       & 21.5       & 79      \\
WISE J1146+4129 & 1549               & 150                &  39.4       & 22.5       & 85      \\
WISE J1603+2745 & 125                & 38                 &  44.0       & 25.5       & 103     \\
WISE J1814+3412 & 157                & 47                 &  37.6       & 23.9       & 76      \\
\hline   
\end{tabular}
\end{table*}

\section{Results and Discussion}
\label{results}

Using the full radio telescope array (including the longest baselines to Sheshan), all four hot DOGs targeted with VLBI remained undetected in the uniformly-weighted dirty image that would provide the highest angular resolution, about 2\,mas~$\times$~9\,mas. This non-detection implies about 1\,mJy\,beam$^{-1}$  brightness upper limit (this corresponds to the $\sim$5$\sigma$ noise level in the dirty images), and allows us to estimate the maximum brightness temperature \citep[e.g.][]{Cond82} of any putative mas-scale compact radio component in these galaxies:
\begin{equation}
\label{Tb}
T_\mathrm{b}<1.22\times 10^{12} (1+z) \frac{S}{\theta_1 \theta_2 \nu^2}\,\mathrm{K},
\end{equation}
where $S$~(Jy) is the flux density upper limit for a non-detected source, $\theta_1$ and $\theta_2$ (mas) are the minor and major axis diameters of the Gaussian restoring beam size (FWHM), and $\nu$~(GHz) is the observing frequency. The value obtained, $T_\mathrm{b}\la$$10^{8}$\,K, indicates that synchrotron self-absorbed ``core'' emission \citep{Blan79} is not present or severely deboosted. Relativistic beaming of radio jets towards the observer is excluded in the nuclei of these galaxies since that would produce brightness temperatures at least two orders of magnitude higher \citep[e.g.][]{Read94}. 

On the other hand, extended emission on angular scales of tens of mas (corresponding to hundreds of pc in linear scale in these galaxies) are detected in all four objects in the naturally-weighted images (Fig.~\ref{images}). The parameters of the circular Gaussian brightness distribution model components fitted in {\sc Difmap}, and their derived monochromatic radio powers and brightness temperatures are given in Table~\ref{models}. The statistical errors of the modelfit parameters are estimated according to \citet{Foma99}. Additional flux density calibration uncertainties are assumed as 5 per cent. The component brightness temperatures are calculated by substituting the circular Gaussian diameter (FWHM) as $\theta_1=\theta_2$ in Eq.~\ref{Tb}. The $T_\mathrm{b}$ values exceed $10^{5}$\,K, the upper limit for galaxies without a central AGN \citep{Cond92}. 

For calculating the radio power,
\begin{equation}
\label{power}
P = 4\pi D_{\rm L}^2 S (1+z)^{-1-\alpha},
\end{equation}
we assumed a steep radio spectrum with spectral index $\alpha=-0.8$ (defined as $S\propto\nu^{\alpha}$, where $S$ is the flux density and $\nu$ the frequency). The two-point spectral index is known only for WISE J1814+3412 \citep{Eise12}. In the absence of flux density mesurements at multiple radio frequencies, we adopt the spectral index value of WISE J1814+3412 for the other three sources as well. In Eq.~\ref{power}, $D_{\rm L}$ denotes the luminosity distance of the source.
 
\begin{table*}
  \centering 
  \caption[]{Parameters of the circular Gaussian model components fitted to the VLBI visibility data, the estimated 1.7-GHz monochromatic powers, and the brightness temperatures for the four hot DOGs.}
  \label{models}
\begin{tabular}{ccccccc}        
\hline                 
Source          & Flux density  & \multicolumn{2}{c}{Relative position}& Diameter     & Power        & Brightness temperature \\
                & $S$ (mJy)     & R.A. (mas)        & Dec. (mas)       & FWHM (mas)   & $P$ ($10^{24}$\,W\,Hz$^{-1}$) & $T_{\rm b}$ ($10^{6}$\,K)\\
\hline                       
WISE J0757+5113 & 0.85$\pm$0.06 &  ...              &  ...             & 27.6$\pm$1.1 & 25.3$\pm$1.8 & 1.6$\pm$0.2 \\
                & 0.22$\pm$0.03 &  $-$51.7$\pm$0.3  &   $-$2.6$\pm$0.3 &  7.2$\pm$0.7 &  6.5$\pm$0.9 & 6.1$\pm$2.1 \\
WISE J1146+4129 & 1.92$\pm$0.11 &  ...              &  ...             & 14.6$\pm$0.3 & 33.3$\pm$1.9 & 11.1$\pm$1.1 \\
                & 1.55$\pm$0.11 &     27.2$\pm$0.6  &  $-$31.0$\pm$0.6 & 26.1$\pm$1.3 & 27.1$\pm$1.9 & 2.8$\pm$0.5 \\
WISE J1603+2745 & 0.23$\pm$0.02 &  ...              &  ...             & 29.9$\pm$2.4 & 10.0$\pm$0.9 & 0.4$\pm$0.1 \\
WISE J1814+3412 & 0.69$\pm$0.06 &  ...              &  ...             & 57.8$\pm$4.1 & 25.6$\pm$2.2 & 0.3$\pm$0.1 \\
                & 0.33$\pm$0.03 & $-$520.0$\pm$1.4  & $-$354.4$\pm$1.4 & 34.1$\pm$2.8 & 12.2$\pm$1.1 & 0.4$\pm$0.1 \\
\hline   
\end{tabular}
\end{table*}

\begin{table*}
  \centering 
  \caption[]{Parameters of the four hot DOGs derived from the EVN observations.}
  \label{sources-derived}
\begin{tabular}{cccccc}        
\hline                 
Source name     & VLBI flux               & VLBI to total           & \multicolumn{2}{c}{VLBI position}           & SFR  \\
                & density$^{\rm a}$ (mJy) & flux density ratio$^{\rm b}$ (\%) & right ascension  & declination    & ($10^3\,M_{\odot}\,{\rm yr}^{-1}$) \\
\hline                       
WISE J0757+5113 & 1.20$\pm$0.08 & 34$\pm$3  & 07$^{\rm h}$ 57$^{\rm m}$ 25\fs0687 & 51\degr 13\arcmin 18\farcs865 &  $<39\pm3$ \\
WISE J1146+4129 & 3.89$\pm$0.18 & 89$\pm$5  & 11$^{\rm h}$ 46$^{\rm m}$ 12\fs8620 & 41\degr 29\arcmin 14\farcs120 &  $<5\pm2$  \\
WISE J1603+2745 & 0.26$\pm$0.02 & 11$\pm$1  & 16$^{\rm h}$ 03$^{\rm m}$ 57\fs3676 & 27\degr 45\arcmin 53\farcs260 &  $<49\pm3$ \\
WISE J1814+3412 & 1.14$\pm$0.08 & 80$\pm$11 & 18$^{\rm h}$ 14$^{\rm m}$ 17\fs3077 & 34\degr 12\arcmin 25\farcs438 &  $<6\pm4$  \\
\hline   
\end{tabular}
\\
$^{\rm a}$ at 1.44\,GHz, estimated from the sum of the component flux densities derived from our 1.7-GHz EVN observations (see Table~\ref{models}, column~2), assuming a spectral index $\alpha=-0.8$, as measured for WISE J1814+3412 by \citet{Eise12};
$^{\rm b}$ the total 1.44-GHz flux densities are given in Table~\ref{sources}, column~7
\end{table*}

The flux densities obtained with VLBI on angular scales of $\sim$10\,mas are smaller than the total flux densities measured at different epochs for these sources (Table~\ref{sources-derived}). This does not formally exclude flux density variability, but does not provide evidence for it either. The lack of variability would be consistent with the absence of compact AGN jets and with the resolved radio structures seen in the EVN images (Fig.~\ref{images}).

The technique of phase-referencing enabled us to accurately derive the positions of the target sources. The coordinates of the phase-reference calibrator objects used for relative astrometry are listed in the International Celestial Reference Frame \citep[ICRF2,][]{Fey15} catalogue{\footnote{\tt{http://hpiers.obspm.fr/icrs-pc/icrf2/icrf2.html}}}. The estimated uncertainties of the hot DOG positions listed in Table~\ref{sources-derived} are within 2~mas in each coordinate. 

\subsection{Notes on Individual Sources}

{\bf WISE J0757+5113.} This source has a radio structure extended to $\sim$100~mas in the east--west direction. Note that the ratio of the FIRST \citep{Whit97} peak brightness (3.19$\pm$0.15~mJy~beam$^{-1}$) and the FIRST integral flux density (3.57$\pm$0.15~mJy, Table~\ref{sources}) is lower than unity, already hinting on the resolved nature of WISE J0757+5113. The sum of the flux densities in the two VLBI components equals to one third of the value expected from the FIRST flux density (Table~\ref{sources-derived}). The flux density resolved out with VLBI (2.37$\pm$0.17~mJy) comes from larger ($\sim 0\farcs1$ to $1\arcsec$) angular scales. This emission is associated with either star-forming or AGN activity, or a combination of both. Below, we show that star formation alone cannot explain the extended radio emission in WISE J0757+5113. The most extreme values of SFR reported e.g. for luminous submillimetre galaxies are $\approx10^{4}$~$M_{\odot}\,{\rm yr}^{-1}$ \citep{Mich10}. In comparison, we can use the conversion relation between the radio power and SFR \citep{Hopk03}, 
\begin{equation}
\label{sfr}
SFR=\frac{P}{1.8 \times 10^{21}}
\end{equation}
to estimate how large the SFR would be if the extended radio emission is entirely attributed to star formation in the host galaxy. Here the SFR is measured in $M_{\odot}\,{\rm yr}^{-1}$ and the 1.4-GHz radio power $P$ in W\,Hz$^{-1}$. We get ($3.9\pm0.3) \times 10^{4}$~$M_{\odot}\,{\rm yr}^{-1}$, which exceeds the highest SFR values found in the literature by at least a factor of $\sim$4. We can therefore conclude that the majority of the radio flux density resolved out with VLBI originates from diffuse AGN-related emission in this source, and not from star-forming activity.   

{\bf WISE J1146+4129.} This is the brightest source in the sample. Although its VLBI structure is also resolved and extended (Fig.~\ref{images}), our EVN observation recovered nearly 90 per cent of its total (FIRST) flux density in two components (Table~\ref{sources-derived}) with a projected linear separation of $\sim$350~pc. These components are rather similar in flux density and angular size, making WISE J1146+4129 a candidate compact symmetric object (CSO). CSOs are sub-galactic ($<1$\,kpc) in size and show double compact lobes \citep[e.g.][]{Phil82,Wilk94,Fant09,AnBa12,An12}. If the nuclear radio source in WISE J1146+4129 is indeed a CSO, the two components detected with VLBI would be the hot spots of a young expanding radio AGN. Future VLBI observations at radio frequencies different from 1.7\,GHz could confirm if their spectrum is steep, and monitoring observations could reveal if the source is really expanding. The kinematic age of CSOs inferred from their expansion velocities is typically in the order of 100--1000~yr. CSOs are known to represent the earliest evolutionary phase of AGN radio sources \citep[e.g.][]{Pola03,AnBa12}.

{\bf WISE J1603+2745.} This source is the most resolved with the EVN in our sample, making it the weakest detection among the four hot DOGs observed. The single 30-mas diameter component is responsible for only $\sim$10 per cent of the total flux density recovered by FIRST (Table~\ref{sources-derived}). The difference in the values of the peak brightness (1.77$\pm$0.14~mJy~beam$^{-1}$) and the integral flux density (2.28$\pm$0.14~mJy) in the FIRST catalogue \citep{Whit97} also suggests that the source is extended on arcsec scale. From the total (FIRST) flux density, 2.02$\pm$0.14~mJy remains undetected with VLBI (Table~\ref{sources-derived}). According to Eq.~\ref{sfr}, this would correspond to an unplausibly high star formation rate of ($4.9\pm0.3) \times 10^{4}$~$M_{\odot}\,{\rm yr}^{-1}$. Like in the case of WISE J0757+5113, the extended radio emission of WISE J1603+2745 cannot be explained by star formation only.   

{\bf WISE J1814+3412.} The 1.7-GHz EVN image of the hot DOG ``prototype'' WISE J1814+3412 \citep{Eise12} shows a symmetric double structure (Fig.~\ref{images}). The angular separation of the components A and B is about 630~mas, which corresponds to 5.1~kpc projected linear separation. Notably, the line connecting the two radio components is roughly perpendicular to the characteristic position angle of the Ly$\alpha$-emitting region ($\sim 140\degr$) that extends to more than 30~kpc in the galaxy \citep{Eise12}. This position angle is also indicated in Fig.~\ref{images} for illustration purposes. Based on the morphology, the component separation, and the radio power ($\ga$$4 \times 10^{25}$~W~Hz$^{-1}$, Table~\ref{models}), the high-resolution radio structure of WISE J1814+3412 can be interpreted as two lobes of a medium-sized symmetric object (MSO). MSOs are double-lobed extragalactic radio sources with lobe separations between 1--15~kpc \citep[e.g.][]{Fant95,Read96}. High-power MSOs represent an early phase of radio-loud AGN development, possibly becoming FR II type sources with edge-brightened lobes \citep{Fana74} at a later stage \citep[e.g.][]{AnBa12}. This happens if their radio AGN activity lasts for a sufficiently long time for the jet to expand into Mpc scales, typically $\la10^{8}$~yr \citep[e.g.][]{Kais07,Anto12}, depending on the jet power and the interstellar environment.

An alternative explanation of the radio structure in WISE J1814+3412 would be a dual AGN \citep[see e.g.][and references therein]{Frey12}. During their formation and growth, galaxies as well as their nuclei naturally go through mergers. If the accretion onto the central supermassive black holes of two merging galaxies is maintained at the same time, dual AGN systems may be observed. However, it is challenging to find unequivocal observational evidence for kpc-scale dual AGN \citep[e.g.][]{Come15}. In the case of WISE J1814+3412, there is currently no other observational indication of duality, apart from the two widely-separated weak VLBI components in Fig.~\ref{images}. However, this object is accompanied by a quasar (5\farcs2; within 42~kpc projected linear distance) and another Lyman-break galaxy (3\farcs8 separation) at the same redshift \citep{Eise12}. It is also suggested that there is an overdensity of serendipitous sources around hot DOGs in general on arcmin scale \citep{Jone14}. 

From our VLBI data, assuming that there is no flux density variability, we find that $\sim 80$ per cent of the total 1.4-GHz radio flux density of WISE J1814+3412 \citep{Eise12} is contained in the two components (Table~\ref{sources-derived}). The remaining $\sim0.3$~mJy (corresponding to $1.1 \times 10^{25}$\,W\,Hz$^{-1}$ power) is resolved out by the interferometer and is explained by emission extended to larger scales. This can in principle be related to startburst activity, or diffuse AGN lobe emission, or both. 
A rough estimation of the upper limit to the SFR is possible if we assume that all radio emission resolved out by the EVN is associated exclusively to star formation. Applying the relation (Eq.~\ref{sfr}) between the SFR and the radio power \citep{Hopk03}, we obtain ${\rm SFR} < (6\pm4) \times 10^3\,M_{\odot}\,{\rm yr}^{-1}$ (Table~\ref{sources-derived}). \citet{Eise12} estimated the SFR for WISE J1814+3412 from the starburst component of their SED fit, assuming that the rest-frame ultraviolet emission is associated with star-forming activity. Considering different initial mass functions and reasonable amounts of extinction, they arrived to a range of possible extinction-corrected SFR values from 180 to 600\,$M_{\odot}\,{\rm yr}^{-1}$ \citep{Eise12}. Our SFR upper limit derived from radio data is about an order of magnitude higher than the SFR estimated by \citet{Eise12} for WISE J1814+3412. Despite the large relative uncertainty of our estimate, it is most likely that the dominant source of the flux density extended up to arcsec angular scales is diffuse AGN-related emission in this object.

From the measured bolometric luminosity and the stellar component of their SED fit, \citet{Eise12} suspect that WISE J1814+3412 does not follow the local black hole mass--bulge mass relation. The apparent deficit of stellar mass suggests that that the bulk of the stars is still to be formed in this galaxy.

\section{Summary and Conclusions}
\label{conclusions}

We performed exploratory high-resolution VLBI observations of a small sample of four hot DOGs (WISE J0757+5113, WISE J1146+4129, WISE J1603+2745, WISE J1814+3412) with the EVN at a single frequency, 1.7~GHz, to reveal the nature of the $\sim$10-mas scale compact radio emission in these sources for the first time. All of them are detected with mJy or sub-mJy radio features extended to projected linear sizes of hundreds to thousands of pc. Because the brightness temperatures implied by the VLBI detections (Table~\ref{models}) are above the $10^{5}$\,K upper limit for ``normal'' galaxies \citep{Cond92}, and because of the high luminosities \citep[e.g.][]{Alex12,Midd13}, our results confirm the presence of an active nucleus in these galaxies, supporting the notion that the radio emission from hot DOGs is at least partly powered by central AGN. Radio observations allow us to peer through the dust that heavily obscures the view to the galactic centres in hot DOGs at most of the other wavebands. There are no very compact (mas-scale or smaller) radio features present in these four sources, indicating that their radio emission is not beamed towards the observer. The unbeamed nature is again consistent with what is known about hot DOGs in general \citep{Tsai15}. Recent 5-GHz VLBI observations of a sample of 90 {\em WISE}-selected heavily obscured radio-loud AGN similar to hot DOGs \citep{Lons15b} also show that much of the radio emission is typically resolved out on mas scales in the 62 objects detected. 

We found that the sum of the VLBI component flux densities is always smaller than the total flux density of the objects investigated. Significant variability is unlikely in these unbeamed sources. Star formation activity in the host galaxy, extended AGN (lobe) emission, or a combination of these two contributions could be responsible for the ``missing'' flux density. The hot DOGs are characterised by intense star formation reaching SFR of thousands of solar masses per year \citep[e.g.][]{Eise12,Tsai15}. With the probable exception of WISE J1146+4129, the sources in our sample show significantly more excess radio emission resolved out with VLBI than that could be explained with star-forming activity alone (Table~\ref{sources-derived}). 

The high-resolution radio structures (Fig.~\ref{images}), the component separations, and radio powers (Table~\ref{models}) fit in the general picture of young radio AGN. In particular, WISE J1146+4129 is a candidate CSO, and the most likely explanation of the 5.1-kpc separation double structure of WISE J1814+3412 is that we see symmetric hot spots of an MSO. The CSO and MSO features are terminal shocks resulting from the interaction of the AGN jet with the ambient interstellar medium in the host galaxy \citep[e.g.][]{Kais97,Kawa06}. These radio sources are known to represent the earliest stages of AGN evolution \citep[e.g.][]{Fant95,Read96,Kawa06,AnBa12}, and the high-power ones with well-confined lobes would grow into FR II galaxies if the AGN activity is maintained for a sufficiently long time \citep{AnBa12}. Future sensitive VLBI observations at multiple frequencies (2.3~GHz and above) could provide information on the spectral properties of the components and verify that their radio spectra are steep, as expected for hot spots. Furthermore, the changing separation of the putative hot spots could be measured with repeated VLBI imaging over the time scale of up to $\sim$10~yr. A definitive proof for young AGN would be to detect the expansion of the sources and to estimate their kinematic age.

It is believed that the starburst and quasar activities, and the growth of supermassive black holes are generally triggered by major mergers of gas-rich galaxies \citep[see e.g.][and references therein]{Hopk08}. Intense starburst activity happens at a later stage of the merger evolution, during the final coalescence of the galaxies. At about the same time, the black holes accrete gas rapidly, although obscured from our view in the optical. Binary black holes spiralling towards final merging are expected at this stage \citep[e.g.][]{Komo03}. It is not clear if the peak of the AGN phase follows \citep[e.g.][]{Hopk12} or precedes the peak star-forming period. If AGN ``turn on'' only after the main starburst episode, then their feedback \citep[e.g.][]{Fabi12} may not affect star formation in the host galaxy so profoundly. Together with other accumulating data on hot DOGs, in infrared and submillimetre bands in particular \citep{Eise12,Wu12,Wu14,Jone14,Asse15,Tsai15}, our results suggest that intense star formation and a young powerful AGN coexist in these objects. However, for WISE J1814+3412, \citet{Eise12} found that the bulk of the stellar mass is still to be built in the host galaxy. Therefore it is likely that the dominant AGN activity actually precedes the major star formation in this object.

DOGs seem to be in a transition period from the dusty starburst-heated to the AGN-dominated phases of the evolution of massive galaxies \citep[e.g.][]{Buss11,Rigu15}, where the AGN feedback is in full action. The same applies to the extremely luminous hot DOGs that form a small sub-class of DOGs. Hot DOGs may signal the end of the DOG phase, and soon become visible quasars \citep[e.g.][]{Brid13,Wu14}. It is promising to identify a well-defined type of objects, the {\em radio-emitting} hot DOGs, where the feedback process can be studied, by connecting them to the youngest radio AGN, as might be implied by our exploratory VLBI observations. High-resolution radio imaging data on a larger sample would be essential to better establish this connection.

\section*{Acknowledgements}

We are grateful to the anonymous referee for valuable comments which helped us to improve the presentation of our results.
The European VLBI Network is a joint facility of independent European, African, Asian, and North American radio astronomy institutes. Scientific results from data presented in this publication are derived from the following EVN project code: EF025.
This work was supported by the Hungarian Scientific Research Fund (OTKA, NN110333), the China Ministry of Science and Technology (2013CB837900), and the China--Hungary Collaboration and Exchange Programme by the International Cooperation Bureau of the Chinese Academy of Sciences.
The research leading to these results has received funding from the European Commission Seventh Framework Programme (FP/2007--2013) under grant agreement No.\,283393 (RadioNet3). 
ZP acknowledges support of the MAGNA project from the International Space Science Institute (ISSI) in Bern, Switzerland. 
TA thanks the Shanghai Rising Star Program.

\label{lastpage}


\begin{thebibliography}{99}

\bibitem[\protect\citeauthoryear{Alexandroff et al.}{2012}]{Alex12} 
Alexandroff R. et al., 2012, MNRAS, 423, 1325

\bibitem[\protect\citeauthoryear{An \& Baan}{2012}]{AnBa12} 
An T., Baan W.A., 2012, ApJ, 760, 77

\bibitem[\protect\citeauthoryear{An et al.}{2012}]{An12} 
An T. et al., 2012, ApJS, 198, 5

\bibitem[\protect\citeauthoryear{Antognini et al.}{2012}]{Anto12}
Antognini J., Bird J., Martini P., 2012, ApJ, 756, 116

\bibitem[\protect\citeauthoryear{Assef et al.}{2015}]{Asse15}
Assef R.J. et al., 2015, ApJ, 804, 27 

\bibitem[\protect\citeauthoryear{Beasley \& Conway}{1995}]{Beas95}
Beasley A.J., Conway J.E., 1995, in Zensus J.A., Diamond P.J., Napier P.J., eds, ASP Conf. Ser. 82, Very Long Baseline Interferometry and the VLBA. Astron. Soc. Pac., San Francisco, p. 327

\bibitem[\protect\citeauthoryear{Blain et al.}{2002}]{Blai02}
Blain A.W., Smail I., Ivison R.J., Kneib J.-P., Frayer D.T., 2002, Phys. Rep., 369, 111

\bibitem[\protect\citeauthoryear{Blandford \& K\"onigl}{1979}]{Blan79} 
Blandford R.D., K\"onigl A., 1979, ApJ, 232, 32

\bibitem[\protect\citeauthoryear{Bridge et al.}{2013}]{Brid13}
Bridge C.R. et al., 2013, ApJ, 769, 91 

\bibitem[\protect\citeauthoryear{Bussmann et al.}{2011}]{Buss11}
Bussmann R.S. et al., 2011, ApJ, 733, 21

\bibitem[\protect\citeauthoryear{Comerford et al.}{2015}]{Come15}
Comerford J.M., Pooley D., Barrows R.S., Greene J.E., Zakamska N.L., Madejski G.M., Cooper M.C., 2015, ApJ, 806, 219 

\bibitem[\protect\citeauthoryear{Condon}{1992}]{Cond92}
Condon J.J., 1992, ARA\&A, 30, 575

\bibitem[\protect\citeauthoryear{Condon et al.}{1982}]{Cond82}
Condon J.J., Condon M.A., Gisler G., Puschell J.J., 1982, ApJ, 252, 102

\bibitem[\protect\citeauthoryear{Cutri et al.}{2012}]{Cutr12}
Cutri R.M. et al., 2012, Explanatory Supplement to the WISE All-Sky Data Release Products, http://wise2.ipac.caltech.edu/docs/release/allsky/expsup

\bibitem[\protect\citeauthoryear{Del Moro et al.}{2013}]{DelM13}
Del Moro A. et al., 2013, A\&A, 549, A59

\bibitem[\protect\citeauthoryear{Dey et al.}{2008}]{Dey08}
Dey A. et al., 2008, ApJ, 677, 943

\bibitem[\protect\citeauthoryear{Diamond}{1995}]{Diam95}
Diamond P.J., 1995, in Zensus J.A., Diamond P.J., Napier P.J., eds, ASP Conf. Ser. 82, Very Long Baseline Interferometry and the VLBA. Astron. Soc. Pac., San Francisco, p. 227

\bibitem[\protect\citeauthoryear{Eisenhardt et al.}{2012}]{Eise12}
Eisenhardt P.R.M. et al., 2012, ApJ, 755, 173 

\bibitem[\protect\citeauthoryear{Fabian}{2012}]{Fabi12} 
Fabian A.C., 2012, ARA\&A, 50, 455

\bibitem[\protect\citeauthoryear{Fanaroff \& Riley}{1974}]{Fana74} 
Fanaroff B.L., Riley J.M., 1974, MNRAS, 167, 31P 

\bibitem[\protect\citeauthoryear{Fanti}{2009}]{Fant09} 
Fanti C., 2009, AN, 330, 120 

\bibitem[\protect\citeauthoryear{Fanti et al.}{1995}]{Fant95} 
Fanti C., Fanti R., Dallacasa D., Schilizzi R.T., Spencer R.E., Stanghellini C., 1995, A\&A, 302, 317 

\bibitem[\protect\citeauthoryear{Fey et al.}{2015}]{Fey15}
Fey A.L. et al. 2015, AJ, 150, 58

\bibitem[\protect\citeauthoryear{Fomalont}{1999}]{Foma99} 
Fomalont E.B., 1999, in Taylor G.B., Carilli C.L., Perley R.A., eds, ASP Conf. Ser. 180, Synthesis Imaging in Radio Astronomy II. Astron. Soc. Pac., San Francisco, p. 301

\bibitem[\protect\citeauthoryear{Frey et al.}{2012}]{Frey12} 
Frey S., Paragi Z., An T., Gab\'anyi K.\'E., 2012, MNRAS, 425, 1185

\bibitem[\protect\citeauthoryear{Heavens et al.}{2004}]{Heav04} 
Heavens A., Panter B., Jimenez R., Dunlop J., 2004, Nat, 428, 625

\bibitem[\protect\citeauthoryear{H\"{o}gbom}{1974}]{Hogb74}
H\"{o}gbom J.A., 1974, A\&AS, 15, 417

\bibitem[\protect\citeauthoryear{Hopkins}{2012}]{Hopk12} 
Hopkins P.E., 2012, MNRAS, 420, L8

\bibitem[\protect\citeauthoryear{Hopkins et al.}{2003}]{Hopk03} 
Hopkins A.M. et al., 2003, ApJ, 599, 971

\bibitem[\protect\citeauthoryear{Hopkins et al.}{2008}]{Hopk08} 
Hopkins P.E., Hernquist L., Cox T.J., Kere\v{s} D., 2008, ApJS, 175, 356

\bibitem[\protect\citeauthoryear{Jones et al.}{2014}]{Jone14}
Jones S.F. et al., 2014, MNRAS, 443, 146 

\bibitem[\protect\citeauthoryear{Kaiser \& Alexander}{1997}]{Kais97} 
Kaiser C.R., Alexander P., 1997, MNRAS, 286, 215

\bibitem[\protect\citeauthoryear{Kaiser \& Best}{2007}]{Kais07} 
Kaiser C.R., Best P.N., 2007, MNRAS, 381, 1548

\bibitem[\protect\citeauthoryear{Kawakatu \& Kino}{2006}]{Kawa06} 
Kawakatu N., Kino M, 2006, MNRAS, 370, 1513
	
\bibitem[\protect\citeauthoryear{Keimpema et al.}{2015}]{Keim15}
Keimpema A. et al., 2015, Exp. Astron., 39, 259

\bibitem[\protect\citeauthoryear{Komossa et al.}{2003}]{Komo03} 
Komossa S., Burwitz V., Hasinger G., Predehl P., Kaastra J.S., Ikebe Y., 2003, ApJ, 599, 971

\bibitem[\protect\citeauthoryear{Lonsdale et al.}{2015a}]{Lons15a}
Lonsdale C.J. et al., 2015a, ApJ, in press (arXiv:1509.00342)

\bibitem[\protect\citeauthoryear{Lonsdale et al.}{2015b}]{Lons15b}
Lonsdale C.J. et al., 2015b, AN, in press 

\bibitem[\protect\citeauthoryear{Micha\l{}owski et al.}{2010}]{Mich10} 
Micha\l{}owski M., Hjorth J., Watson D., 2010, A\&A, 514, A67

\bibitem[\protect\citeauthoryear{Middelberg et al.}{2013}]{Midd13} 
Middelberg E. et al., 2013, A\&A, 551, A97

\bibitem[\protect\citeauthoryear{Narayanan et al.}{2010}]{Nara10}
Narayanan D. et al., 2010, MNRAS, 407, 1701

\bibitem[\protect\citeauthoryear{Phillips \& Mutel}{1982}]{Phil82} 
Phillips R.B., Mutel R.L., 1982, A\&A, 106, 21 

\bibitem[\protect\citeauthoryear{Piconcelli et al.}{2015}]{Pico15}
Piconcelli E. et al., 2015, A\&A, 574, L9

\bibitem[\protect\citeauthoryear{Polatidis \& Conway}{2003}]{Pola03} 
Polatidis A.G., Conway J.E., 2003, PASA, 20, 69 

\bibitem[\protect\citeauthoryear{Readhead}{1994}]{Read94}
Readhead A.C.S., 1994, ApJ, 426, 51 

\bibitem[\protect\citeauthoryear{Readhead et al.}{1996}]{Read96}
Readhead A.C.S., Taylor G.B., Pearson T.J., Wilkinson P.N., 1996, ApJ, 460, 634 

\bibitem[\protect\citeauthoryear{Riguccini et al.}{2015}]{Rigu15}
Riguccini L. et al., 2015, MNRAS, 452, 470

\bibitem[\protect\citeauthoryear{Rowan-Robinson}{2000}]{Rowa00}
Rowan-Robinson M., 2000, MNRAS, 316, 885 

\bibitem[\protect\citeauthoryear{Schwab \& Cotton}{1983}]{Schw83}
Schwab F.R., Cotton W.D., 1983, AJ, 88, 688

\bibitem[\protect\citeauthoryear{Shepherd et al.}{1994}]{Shep94}
Shepherd M.C., Pearson T.J., Taylor G.B., 1994, BAAS, 26, 987

\bibitem[\protect\citeauthoryear{Stern et al.}{2014}]{Ster14}
Stern D. et al., 2014, ApJ, 794, 102 

\bibitem[\protect\citeauthoryear{Szomoru}{2008}]{Szom08}
Szomoru A., 2008, Proceedings of Science, PoS(IX EVN Symposium)040

\bibitem[\protect\citeauthoryear{Tsai et al.}{2015}]{Tsai15}
Tsai C.-W. et al., 2015, ApJ, 805, 90 

\bibitem[\protect\citeauthoryear{White et al.}{1997}]{Whit97}
White R.L., Becker R.H., Helfand D.J., Gregg M.D., 1997, ApJ, 475, 479 

\bibitem[\protect\citeauthoryear{Wilkinson et al.}{1994}]{Wilk94} 
Wilkinson P.N., Polatidis A.G., Readhead A.C.S., Xu W., Pearson T.J., 1994, ApJ, 432, L87 

\bibitem[\protect\citeauthoryear{Wright}{2006}]{Wrig06}
Wright E.L., 2006, PASP, 118, 1711 

\bibitem[\protect\citeauthoryear{Wright}{2010}]{Wrig10}
Wright E.L. et al., 2010, AJ, 140, 1868 

\bibitem[\protect\citeauthoryear{Wu et al.}{2012}]{Wu12}
Wu J. et al., 2012, ApJ, 756, 96 

\bibitem[\protect\citeauthoryear{Wu et al.}{2014}]{Wu14}
Wu J. et al., 2014, ApJ, 793, 8 


\end{thebibliography}
\end{document}